\documentclass[aip,apl,reprint,a4paper]{revtex4-1}
\usepackage{graphicx}

\begin{document}

\title{Nanotube-based scanning rotational microscope}

\date{\today}

\author{Andrey M. Popov}
\email{am-popov@isan.troitsk.ru}
\affiliation{Institute of Spectroscopy of Russian Academy of Sciences, Fizicheskaya Street 5, Troitsk 142190, Moscow Region, Russia}
\author{Irina V. Lebedeva}
\email[Author to whom correspondence should be addressed. Electronic mail: ]{lebedeva@kintechlab.com}
\affiliation{Kintech Lab Ltd., Kurchatov Square 1, Moscow 123182, Russia}
\author{Andrey A. Knizhnik}
\email{knizhnik@kintechlab.com}
\affiliation{Kintech Lab Ltd., Kurchatov Square 1, Moscow 123182, Russia}
\affiliation{National Research Centre ``Kurchatov Institute'', Kurchatov Square 1, Moscow 123182, Russia}

\begin{abstract}
A scheme of the scanning rotational microscope is designed. This scheme is based on using carbon nanotubes simultaneously as a probe
tip and as a bolt/nut pair which converts translational displacements of two piezo actuators into pure rotation of the probe tip. First-principles
calculations of the interaction energy between movable and rotational parts of the microscope confirms the capability for its operation. The scanning
rotational microscope with a chemically functionalized nanotube-based tip can be used to study how the interaction between individual molecules
or a molecule and a surface depends on their relative orientation.
\end{abstract}

\maketitle

The high aspect ratio of carbon nanotubes and ability to buckle elastically make them perfectly suitable for application as probe tips in
scanning probe microscopy.\cite{dai96} Moreover, it was shown that functionalization of carbon nanotubes gives the possibility to use them as
tips of chemical probes.\cite{wong98} A wide set of functionalized or coated carbon nanotube-based probe tips have been used in various areas of
physics, chemistry and biology.\cite{wong99,patil04,deng04,esplandiu04} At the same time, the possibility of relative motion of walls in multi-walled
carbon nanotubes \cite{cumings00,kis06} allows to use these walls as movable elements of nanoelectromechanical systems (see
Ref.~\onlinecite{lozovik07} for a review). Nanomotors in which walls of multi-walled nanotubes play roles of a shaft and a bush
\cite{fennimore03,bourlon04,barreiro08,subramanian07} and memory cells based on relative sliding of the walls along the nanotube axis
\cite{deshpande06,subramanian10} were recently implemented. It was proposed that double-walled nanotubes (DWNTs) can operate as  bolt/nut pairs
\cite{saito01,lozovik03,lozovik03a,barreiro08} and thus can be used in nanoactuators in which a force directed along the nanotube axis leads to relative rotation of the walls.\cite{popov07}

Combining applications of carbon nanotubes in scanning probe microscopy and nanoelectromechanical
systems based on relative motion of nanotube walls, fundamentally new nanodevices can be developed. For example, a DWNT attached to a cantilever of the atomic force microscope was proposed to be used for thermal nanolithography with improved spatial resolution.\cite{popescu09} In the present Letter we suggest a concept of the nanotube-based scanning
rotational microscope (SRM) combining the following applications of carbon nanotubes: 1) use of carbon nanotubes as chemical probe tips
in scanning probe microscopy, 2) possibility of DWNTs to operate as a bolt/nut pair. A principal scheme of the SRM head which allows to
realize pure rotation of the probe tip relative the object of investigation is designed. First-principles calculations of the interaction energy between rotational and movable parts of the SRM confirm the capability for its operation. Possible
applications of the SRM for investigation of the dependence of interaction between individual molecules or a molecule and a surface on their relative
orientation are discussed.

\begin{figure}[b!]
\includegraphics[width=85mm]{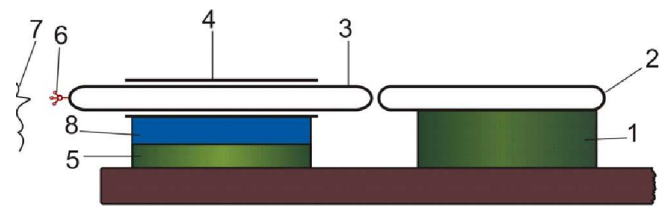}
\caption{\label{Figure1} The scheme of the scanning rotational microscope (SRM) head with a chemically functionalized tip. A piezo actuator (1) is
applied to move a single-walled nanotube (2) and thus induce screw motion of the inner wall (3) of the double-walled nanotube used as a bolt/nut pair
relative to the outer wall (4). A piezo actuator (5) is used to prevent the translational motion of the inner wall (3) with a chemically
attached molecule (6) relative the surface or nanoobject of investigation (7). An electrode (8) is used to apply voltage and measure the tunneling current between
the molecule (6) and the nanoobject (7).}
\end{figure}

The proposed scheme of the SRM head which is based on two carbon nanotubes is presented in Fig.~\ref{Figure1}. In this scheme, a molecule (6) is
chemically attached to a tip of the probe based on the movable inner wall (3) of the DWNT. To study how the interaction between
this molecule and another molecule or a nanoobject (7) depends on their relative orientation pure rotation of the molecule attached to the probe tip relative the nanoobject of investigation with no simultaneous translational displacement should be realized. In the case of the DWNT representing a bolt/nut pair, the
translational motion of the inner wall relative the outer wall (4) is followed by their relative rotation. Thus simultaneous rotational and translational motion of the inner wall
relative to the outer wall can be induced by pushing or pulling the inner wall with a piezo actuator (1). Pushing the DWNT as a whole in the opposite direction using another piezo actuator (5) attached to the outer wall of the DWNT, the translational motion of the inner wall can be
compensated and thus the pure rotation of the inner wall relative to the object of investigation can be achieved. The successful operation of the proposed SRM is possible
if the movable inner wall of the DWNT (3) which is used as the probe can rotate freely relative to the nanoobject (2) which transfers the translational motion from the piezo actuator (5) to the inner wall inducing its simultaneous translational and rotational motion relative to the outer wall. Otherwise the twist-off of the nanotube-based bolt/nut pair would take place. We show below that the second nanotube attached to
a piezo actuator can serve to as this nanoobject (2) inducing the relative screw motion of the DWNT walls. The tunneling current between the molecule (6) and the nanoobject (7) can be measured with an electrode (8) attached to the outer wall of the DWNT.

To confirm that the proposed scheme allows the SRM head to operate we have performed the following calculations. First we have calculated the
interwall interaction energy of a DWNT representing a bolt/nut pair as a function of the relative position of the walls. Second we have obtained
the potential relief of interaction energy between capped ends of coaxial single-walled nanotubes and have shown that free relative rotation of these nanotubes about their
common axis takes place.

In the general case, a nanotube wall has a helical symmetry. Therefore, relative screw motion of the walls can be expected for majority of DWNTs.
However, from symmetry considerations, it follows that the corrugation of the potential relief of interwall interaction energy in DWNTs with perfect chiral walls is too small, complicating application of such nanotubes in nanodevices as bolt/nut pairs.\cite{lozovik05,lozovik06} Recently it was shown that an atomic
scale design of the wall structure makes it possible to produce nanotubes with characteristics of the potential energy relief suitable for their use as the bolt/nut pairs.\cite{belikov04,lozovik05,lozovik06} Namely, it was proposed to create artificial
defects in one of the walls of a nanotube with commensurate chiral walls at an identical position in many unit cells of the nanotube. In this case,
any barrier to relative motion of the nanotube walls is proportional to the number of unit cells with the defects. Thus,
such an atomic scale design of the DWNT structure allows to obtain the bolt/nut pairs with desirable (sufficiently great) values of the barrier to
twist-off. As shown in papers,\cite{lozovik05,lozovik06} qualitative characteristics of the relative screw motion of the walls do not depend on the type of the defects created.

To study the possibility of relative screw motion of nanotube walls, the dependence of interwall interaction energy $U$ of two
neighbouring nanotube walls on their relative position should be calculated. It is convenient to visualize the potential relief of interwall
interaction energy $U(x, \phi)$ as a map plotted on a cylindrical surface, where $x$ is the relative displacement of the walls along the nanotube axis and
$\phi$ is the angle of relative rotation of the walls about the nanotube axis. In principle, a DWNT can operate as the bolt/nut pair if the
potential energy relief has valleys directed along a helical line by analogy with a thread on a lateral bolt surface. Quantitative characteristics of
this thread include potential barriers $E_1$ and $E_2$ to relative screw motion of the DWNT walls along the thread line and across it (i.e.
$E_2$ is the barrier to twist-off or the thread depth). The quality of the thread can be characterized by the
ratio of these barriers, the relative depth of the thread $\beta=E_2/E_1$.\cite{lozovik03,lozovik03a} Evidently, a high value of the relative depth
of the thread $\beta$ implies that there is a wide interval of forces which are sufficient to induce the relative screw motion of the nanotube walls but are too low to result in the twist-off.

\begin{figure}[t!]
\includegraphics{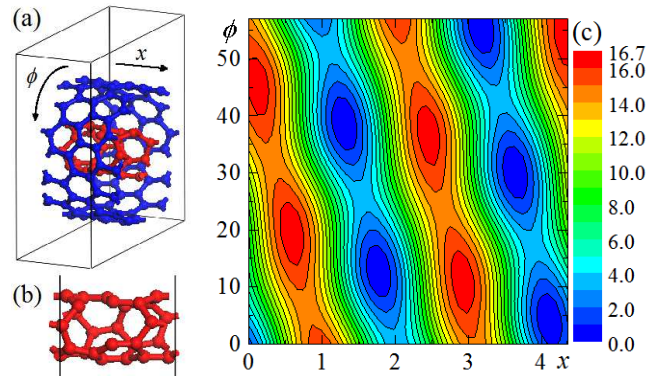}
\caption{\label{Figure2} (Color online) (a) Model cell with one unit cell of the (4,1)@(12,3) DWNT. (b) The inner (4,1) DWNT with a single vacancy in each unit cell. Carbon atoms of the inner and outer walls are colored in red and blue, respectively. The periodic boundary condition is
applied along the x-axis. (c) Calcualted potential relief of interwall interaction energy $U$ (in meV per nanotube unit cell) as a function
of the relative displacement $x$ (in \AA) of the walls along the nanotube axis and the angle $\phi$ (in degrees) of the relative rotation of the
walls about the axis for the (4,1)@(12,3) DWNT with a single vacancy in each unit cell of the inner (4,1) wall. The energy is given relative to the
global energy minimum. 
}
\end{figure}

The potential reliefs of interwall interaction energy in DWNTs with periodic structural defects have been previously studied only using  empirical
potentials.\cite{lozovik05, lozovik06, belikov04} In the present Letter we confirm the conclusion that such DWNTs can be used as bolt/nut pairs by first-principles calculations. As an
example, we consider a (4,1)@(12,3) nanotube with a single vacancy defect in each unit cell of the inner (4,1) wall (Fig.~\ref{Figure2}). Spin-unrestricted density
functional theory calculations are performed using the VASP code \cite{VASP} with the functional of Perdew, Burke, and Ernzerhof\cite{PBE}. The dispersion correction is calculated using the DFT-D2  method of Grimme.\cite{grimme06} The
basis set consists of plane waves with the maximum kinetic energy of 700 eV. The interaction of valence electrons with atomic cores is described
using the projector augmented-wave method (PAW). \cite{PAW} A second-order Methfessel-Paxton smearing \cite{MPsmearing} with a width of 0.05 eV is
applied. The model cell comprising one unit cell of the nanotube 6.528~\AA~x~18~\AA~x~18~\AA~  is considered. The periodic boundary condition is
applied along the x-axis. Integration over the Brillouin zone is performed using the Monkhorst-Pack method \cite{MPmethod} with an 14~x~1~x~1~ k-point
sampling. The structures of the inner wall with the vacancy and of the outer wall are separately geometrically optimized until the residual force acting on each atom is less than $10^{-3}$~eV/\AA. Then they are rigidly shifted and rotated
relative to each other. Account of deformation of nanotube  walls was proved to be inessential for the shape of potential energy relief both for the
interwall interaction energy of DWNTs\cite{kolmogorov00} and the intershell interaction energy of carbon nanoparticles.\cite{lozovik00,lozovik02}

The calculated potential relief of interwall interaction energy for the (4,1)@(12,3)DWNT with a single vacancy in each unit cell of the inner wall has a thread-like shape (Fig.~\ref{Figure2}). The potential barriers to relative screw motion of the DWNT
walls along and across the thread line are found to be $E_1=3.25$ meV and $E_2=14.6$ meV per unit cell of the (4,1)@(12,3) DWNT,
respectively. The convergence tests show that these values are accurate to less than 5\%. Contibutions of the van der Waals attraction to the barriers $E_1$ and $E_2$ are revealed to be as small as 13\% and 3\%, respectively. This is in agreement with previous calculations which showed that the van der Waals attraction provides a minor correction to potential reliefs of interaction energy between graphene layers\cite{lebedeva11} and between polycyclic aromatic molecules and a graphene flake.\cite{ershova10}
The relative depth of the thread is estimated to be $\beta \approx 4.5$. This is sufficient for the use of the (4,1)@(12,3) DWNT with
periodic vacancies as a bolt/nut pair. 

It should be noted that the calculated potential energy relief can be roughly fitted with the following simple expression containing only the first Fourier components \cite{kuznetsov12}

\begin{eqnarray} 
\label{approx}
\nonumber
U(x, \phi) = U_0 + \frac{E_2}{2} \left[ 1 - \cos \left( k x \cos \alpha + k R \phi \sin \alpha \right) \right]+
\\
\nonumber
\frac{E_1}{2} \left[ 1 - \cos \left( k x \cos \left( \frac{\pi}{3} - \alpha \right) - k R \phi \sin \left( \frac{\pi}{3} - \alpha \right)
\right) \right],
\\
\end{eqnarray}
where $U_0$ is the energy in the energy minima, $k=4 \pi/(3l)$, $l=1.42$~\AA~is the bond length in carbon nanotubes, $R=5.4$~\AA~is the radius of the perfect (12,3) wall and $\alpha=11^\circ
$ is the chiral angle of the walls. The root-mean-square deviation of approximation (\ref{approx}) with the estimated values of the barriers $E_1$ and $E_2$ given above from the calculated potential energy relief is 0.5~meV per unit cell of the DWNT, which is much smaller than the magnitude of corrugation of the potential energy relief about 16.7~meV per unit cell of the DWNT. The first Fourier components were previously shown to be sufficient for description of the potential reliefs of interwall interaction energy in DWNTs with commensurate nonchiral
walls\cite{belikov04,bichoutskaia09} and of interlayer interaction energy in bilayer graphene.\cite{lebedeva11}

\begin{figure}[t!]
\includegraphics{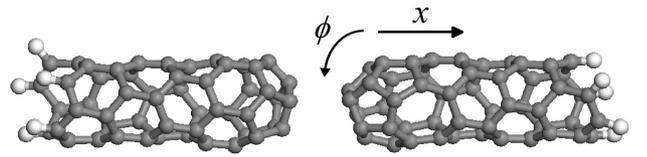}
\caption{\label{Figure3} Two (4,1) nanotubes interacting via their caps. Carbon and hydrogen atoms are colored in gray and white, respectively. }
\end{figure}

As discussed above, not only deepness of the thread but also free relative rotation of the inner wall of the first DWNT
and of the second single-walled nanotube interacting via their caps (Fig.~\ref{Figure1}) are necessary for operation of the proposed SRM. Namely, the barrier in the dependence of the
interaction energy between these caps on the angle of relative rotation of the nanotubes about their common axis should be considerably less that
the barrier to twist-off of the DWNT-based bolt/nut pair. To investigate the relation between these two barriers for the (4,1)@(12,3) DWNT interacting
with another (4,1) nanotube we have calculated the potential relief of interaction energy between two capped (4,1) nanotubes. The cap for the (4,1)
nanotube is built of 1 heptagon and 7 pentagons (Fig.~\ref{Figure3}). The other end of the
nanotube of 12~\AA~length is terminated with hydrogen atoms. The capped (4,1) nanotube obtained in this way is geometrically optimized. Its inversion in a point on the nanotube axis yeilds the second capped (4,1) nanotube. Then one of these (4,1) nanotubes with the interacting caps is rigidly shifted and rotated
around their common axis (Fig.~\ref{Figure3}). In these calculations, the size of the model cell is 40~\AA~x~14~\AA~x~14~\AA. A single G-point is used for integration over
the Brillouin zone. The
basis set consists of plane waves with the maximum kinetic energy of 500 eV. 

Our calculations show that the energy minimum for the interaction between the coaxial capped (4,1) nanotubes is reached at the distance of 3.9~\AA~between
the nanotubes. The barrier to relative rotation of the nanotubes around their common axis is found to be 62 meV. It is seen that this barrier should
be more than an order of magnitude smaller than the barrier $E_2$ to twist-off of the bolt/nut pair based on the (4,1)@(12,3) DWNT with a hundred
 vacancies in the inner wall. Thus our calculations demonstrate that the translational motion of the piezo-actuator can be converted to the relative screw motion of the nanotube-based bolt/nut pair (Fig.~\ref{Figure1})
 by the interaction between the caps of coaxial nanotubes.

Since the pioneering experiment \cite{lyo91} a wealth of experience was accumulated in recent twenty years in controlled manipulation (both deposition and creation of a vacancy) of individual atoms on a surface using scanning probe microscopy (see Refs.~\onlinecite{tseng08, custance09} for reviews). A considerable progress in manipulation with nanoobjects has made it possible not only to create individual nanomotors \cite{fennimore03, bourlon04, barreiro08} and memory cells \cite{deshpande06} based on relative motion of nanotubes walls but also to elaborate methods of their batch fabrication.\cite{subramanian07, subramanian10} All these give us a cause for optimism that the proposed nanotube-based SMR will be implemented in the near future.

To summarize, in the present Letter we designed the scheme of the SRM head. The proposed SRM is based on two carbon
nanotubes. The first nanotube is used simultaneously as a probe tip and as a bolt/nut pair which converts translational displacements of two piezo
actuators into pure rotation of the probe tip. The second single-walled nanotube is used to induce relative screw motion in the bolt/nut pair.

The first-principles calculations of the thread-like potential relief of interwall interaction energy for DWNTs with commensurate chiral walls containing periodic
defects examplified
by the (4,1)@(12,3) DWNT with vacancies showed that such nanotubes can be easy-to-use bolt/nut pairs. The
possibility to use a single-walled nanotube to induce the relative screw motion in the nanotube-based bolt/nut pair was confirmed by the calculations
of the potential relief of interaction energy between the coaxial capped (4,1) nanotubes.

Let us discuss possible applications of the SRM. Functionalization of the SRM probe tip with different chemical groups
allows to use the carbon nanotube-based probe tip as a chemical probe and thus opens up possibilities to study how interaction (and even probability of chemical reactions)
between individual molecules or a molecule and a surface depends on their relative orientation. Any atomic scale translational displacement of the SRM head as a whole can be realized in the same way as the
translational displacement of heads of scanning probe microscopes. Thus, the proposed SRM can be used to study
dependences of interaction between individual molecules on their relative position and orientation simultaneously. The SRM is particularly promising for investigation of interaction
between individual macromolecules such as an antigen and an antibody, where atomic force microscopy is widely used.\cite{leckband00,allison02}

\begin{acknowledgments}
This work has been partially supported by the Russian Foundation of Basic Research (grants 11-02-00604-a and 12-02-90041-Bel). The calculations are performed on the SKIF MSU Chebyshev supercomputer, on the MVS-100K supercomputer at the Joint Supercomputer Center of the Russian Academy of Sciences and on the Multipurpose Computing Complex NRC ``Kurchatov Institute''.
\end{acknowledgments}

\end{document}